\documentclass[sigconf]{acmart}

\usepackage{booktabs} 
\usepackage{multirow}
\usepackage{makecell}

\usepackage{subcaption}

\copyrightyear{2020}
\acmYear{2020}
\setcopyright{acmcopyright}\acmConference[JCDL '20]{Proceedings of the ACM/IEEE Joint Conference on Digital Libraries in 2020}{August 1--5, 2020}{Virtual Event, China}
\acmBooktitle{Proceedings of the ACM/IEEE Joint Conference on Digital Libraries in 2020 (JCDL '20), August 1--5, 2020, Virtual Event, China}
\acmPrice{15.00}
\acmDOI{10.1145/3383583.3398540}
\acmISBN{978-1-4503-7585-6/20/06}

\begin{document}
\fancyhead{}                
\title{Identifying Documents In-Scope of a Collection \\ from Web Archives}

\author{Krutarth Patel}
\email{kipatel@ksu.edu}
\affiliation{%
  \institution{Kansas State University}
  \city{Computer Science}
}

\author{Cornelia Caragea}
\email{cornelia@uic.edu}
\affiliation{%
  \institution{University of Illinois at Chicago}
  \city{Computer Science}
  }
  
\author{Mark E. Phillips}
\email{Mark.Phillips@unt.edu}
\affiliation{%
  \institution{University of North Texas}
  \city{UNT Libraries}
  }

\author{Nathaniel T. Fox}
\email{Nathan.Fox@unt.edu}
\affiliation{%
  \institution{University of North Texas}
  \city{UNT Libraries}
}

\begin{abstract}

Web archive data usually contains high-quality documents that are very useful for creating specialized collections of documents, e.g., scientific digital libraries and repositories of 
technical reports. In doing so, there is a substantial need for automatic approaches that can distinguish the documents of interest for a collection out of the huge number of documents collected by web archiving institutions. In this paper, we explore different learning models and feature representations to determine the best performing ones for identifying the documents of interest from the web archived data. Specifically, we study both machine learning and deep learning models and ``bag of words'' (BoW) features extracted from the entire document or from specific portions of the document, as well as structural features that capture the structure of documents. We focus our evaluation on three datasets that we created from three different Web archives. Our experimental results show that the BoW classifiers that focus only on specific portions of the documents (rather than the full text) outperform all compared methods on all three datasets. 

\end{abstract}


\maketitle

\section{Introduction}
A growing number of research libraries, museums, and archives around the world are embracing web archiving as a mechanism to collect born-digital material made available via the web. Between the membership of the International Internet Preservation Consortium, which has 55 member institutions \cite{iipc2017membership}, and the Internet Archive's Archive-It web archiving platform with its 529 collecting organizations \cite{internetarchive2017}, there are 
hundreds of institutions currently engaged in building collections with web archiving tools. The amount of data that these web archiving initiatives generate is typically at levels that dwarf traditional digital library collections. As an example, in a recent impromptu analysis, Jefferson Bailey of the Internet Archive noted that there were 1.6 Billion PDF files in the Global Wayback Machine \cite{bailey2017pdfs}. If just 1\% of these PDFs are of interest for collection organizations, that would result in a collection larger than the 15 million volumes in HathiTrust \cite{hathitrustNDvolumes}. 

Interestingly, while the number of web archiving institutions and collections increased in recent years, the technologies needed to 
extract high-quality, content-rich PDF documents from the web archives in order to add them to their existing collections and repository systems have not improved significantly over the years. 

Our research is aimed at understanding how well machine learning and deep learning models can be employed to provide assistance to collection maintainers who are seeking to classify the PDF documents from their web archives into being within scope for a given collection or collection policy or out of scope. By identifying and extracting these documents, institutions will improve their ability to provide meaningful access to collections of materials harvested from the web that are complementary, but oftentimes more desirable than traditional web archives. At the same time, building specialized collections from web archives shows usage of web archives beyond just replaying the web from the past. Our research focus is on three different use cases that have been identified for the reuse of web archives and include populating an institutional repository from a web archive of a university domain, the identification of state publications from a web archive of a state government, and the extraction of technical reports from a large federal agency. These three use cases were chosen because they have broad applicability and cover different Web archive domains.

Precisely, in this paper, we explore and contrast different learning models and types of features to determine the best performing ones for identifying the documents of interest that are in-scope of a given collection. Our study includes both an exploration of traditional machine learning models in conjunction with either a ``bag of words'' representation of text or structural features that capture the characteristics of documents, as well as an exploration of Convolutional Neural Networks (CNN). 
The ``bag of words'' (BoW) or \textit{tf-idf} representation is commonly used for text classification problems~\cite{caropreso2001learner,Sebastiani2002}. Structural features designed based on the structure of a document have been successfully used for document type classification \cite{caragea2016document}.
Moreover, for text classification tasks, Convolutional Neural Networks 
are also extensively used in conjunction with word embeddings and achieve remarkable results~\cite{goldberg2016primer,johnson2014effective,kim2014convolutional}. 

In web archived collections, usually different types of documents have a different textual structure and cover different topics. The beginning and the end portions of a document generally contain useful and sufficient information (either structural or topical) for deciding if a document is in scope of a collection or not. As an example, consider a scholarly works repository. Research articles usually contain the abstract and introduction in the beginning, and conclusion and references in the end. The text of these sections and their position in the document are often enough to determine the classification of documents \cite{caragea2016document}. Being on the proper subject (or in scope of a collection) can also be inferred from the beginning and the end portions of a document, with some works using only the title and abstract for classification \cite{DBLP:conf/icml/LuG03,CarageaSKCM11}. These aspects motivated us to explore features extracted from different portions of documents. 
%
%
To this end, we consider the task of finding documents being in-scope of a collection as a binary classification task. In our work, we experiment with bag of words (``BoW'') by using text from the entire document as well as by focusing only on specific portions of the document (i.e., the beginning and the end part of the document). Although structural features were originally designed for document type classification, we used these features for our binary classification task. We also experiment with a CNN classifier that exploits pre-trained word embeddings. 

In summary, our contributions are as follows:
\begin{itemize}
\item We built three datasets from three different web archives collected by the UNT libraries, each covering different domains: UNT.edu, Texas.gov, and USDA.gov. Each dataset contains the PDF document along with the label indicating whether a document is in scope of a collection or not. We  will make these datasets available to further research in this area.
\item We show that BoW classifiers that use only some portion of the documents outperform BoW classifiers that use full text from the entire content of a document, the structural features based classifiers, and the CNN classifier. 
\item We also show that feature selection using information gain improves the performance of the BoW classifiers and structural features based classifiers, and present a discussion on the most informative features for each collection.
\end{itemize}


\section{Related Work}
\label{sec:rel_work}

{\bf \em Web Archiving.} Web archiving as a method for collecting content has been conducted by libraries and archives since the mid-1990's. The most known web archiving is operated by the Internet Archive who began harvesting content in 1996. Other institutions throughout the world have also been involved in archiving the web based on their local collection context whether it is based on a subject or as part of a national collecting mandate such as with national libraries across the world.  
While the initial focus of these collections was to preserve the web as it exists in time, there were subsequent possibilities to leverage web archives to improve access to resources that have been collected, for example, after the collaborative harvest of the federal government domain by institutions around the United States for the 2008 End of Term Web Archive whose goal was to document the transitions from the Bush administration to the Obama administration. After the successful collection of over 16TB of web content, Phillips and Murray \cite{Phillips2013pdf} analyzed the 4.5M unique PDFs found in the collection to better understand their makeup. Jacobs \cite{jacobs2014govdocs} articulated the value and importance of web-published documents from the federal government that are often found in web archives. Nwala et al.~\cite{nwala2018bootstrapping} studied bootstrapping of the web archive collections from the social media 
and showed that sources such as Reddit, Twitter, and Wikipedia can produce collections that are similar to expert generated collections (i.e., Archive-It collections). 
Alam et al.~\cite{alam2015improving} proposed an approach to index raster images of dictionary pages and  built a Web application that supports word indexes in various languages 
with multiple dictionaries. 
Alam et al.~\cite{alam2016web} used CDX summarization for web archive profiling, whereas AlNoamany et al.~\cite{alnoamany2017generating} proposed the Dark and Stormy Archive (DSA) framework for summarizing the holdings
of these collections and arranging them into a chronological order. 
Aturban et al.~\cite{aturban2019archive} proposed two approaches to to establish and check fixity of archived resources.
More recently, the Library of Congress \cite{dooley2019webarchives} analyzed its web archiving holdings and identified 42,188,995 unique PDF documents in its holdings. These initiatives show interest in analyzing the PDF documents from the web as being of interest to digital libraries. Thus, there is a need however for effective and efficient tools and techniques to help filter or sort the desirable PDF content from the less desirable content based on existing collections or collection development policies. 
%
We specifically address this with our research agenda and formulate the problem of classifying the PDF documents from the web archive collection into being of scope for a given collection or being out of scope. We use both traditional machine learning and deep learning models. Below we discuss related works on both of these lines of research. 
	

{\bf \em Traditional Text Classification.} Text classification is a well-studied problem. The {\em BoW} ({\em binary}, {\em tf}, or {\em tf-idf}) representations are commonly used as input to machine learning classifiers, e.g., Support Vector Machine \cite{Joachims1998} and Na\"ive Bayes Multinomial  \cite{McCallum98} for text classification. Feature selection is often applied to these representations to remove irrelevant or redundant features \cite{Forman2003,Dumais1998}. In the context of digital libraries, the classes for text classification are often document topics, e.g., papers classified as belonging to ``machine learning'' or ``information retrieval'' \cite{lu03icml}. 
Structural features that capture the structural characteristics of documents are also used for the classification of documents in digital libraries \cite{caragea2016document}. 
Comprehensive reviews of the feature representations, methods, and results on various text classification problems are provided by Sebastiani~\cite{Sebastiani2002} and Manning~\cite{Manning2008}.
Craven and Cumlien~\cite{craven1999constructing} classified bio-medical articles using the Naive Bayes classifier. Kodakateri Pudhiyaveetil et al.~\cite{kodakateri2009conceptual} used the k-NN classifier to classify computer science papers into 268 different categories based on the ACM classification tree. Other authors \cite{zhou2016classifying,HaCohen-Kerner13} 
experimented with different classification methods such as unigram, bigram, and Sentence2Vec~\cite{le2014distributed} to identify the best classification method for classifying academic papers using the entire content of the scholarly documents.

{\bf \em Deep Learning.} Deep learning models have achieved remarkable results in many NLP and text classification problems~\cite{bengio2003neural,collobert2008unified,collobert2011natural,mikolov2013distributed,kalchbrenner2014convolutional,goldberg2016primer}. Most of the works for text classification with deep learning methods have involved word embeddings. Among different deep learning architectures, convolutional neural network (CNN), recurrent neural network (RNN), and their variations are the most popular architectures for text applications.
Kalchbrenner et al.~\cite{kalchbrenner2014convolutional} proposed a deep learning architecture with multiple convolution layers that uses word embeddings initialized with random vectors. Zhang et al.~\cite{zhang2015character} used encoded characters (``one-hot'' encoding) as an input to the deep learning architecture with multiple convolution layers. They proposed a 9-layer deep network with 6 convolutional layers and 3 fully-connected layers. Kim~\cite{kim2014convolutional} used a single layer of CNN after extracting word embeddings for tokens in the input sequence. The author experimented with several variants of word embeddings, i.e., randomly initialized word vectors later tuned for a specific task, fixed pre-trained vectors, pre-trained vectors later tuned for a specific task, and a combination of the two sets of word vectors.  
Yin et al.~\cite{yin2016multichannel} used the combination of diverse versions of pre-trained word embeddings followed by a CNN and a fully connected layer for the sentence classification problem. 

In our work, we compare the performance of the BoW classifier that use the entire text of documents, with those that use only some portions of the documents, with structural features based classifiers proposed by Caragea et al.~\cite{caragea2016document}, and with a CNN classifier that uses pre-trained word embeddings (which is more efficient that an RNN-based classifier). 
We also experiment with top-N selected features, ranked using the information gain feature selection method for the BoW and structural features extracted from the entire documents.

\section{Data}
\label{sec:data}

For this research, we constructed datasets from three web archives collected by the UNT Libraries. For each of the datasets we extracted all PDF documents within each of the web archives. 
Next, we randomly sampled 2,000 PDFs from each collection that we used as the basis for our labeled datasets. Each of the three sets of 2,000 PDF documents were then labeled in scope and out of scope by subject matter experts who are responsible for collecting publications from the web for their collections. Each dataset includes PDF files along with their labels (in scope/out of scope or relevant/irrelevant). 
Further description of the datasets is provided 
below.

\subsection{UNT.edu dataset}
The first dataset was created from the UNT Scholarly Works web archive of the unt.edu domain. This archive was created in May 2017 as part of a bi-yearly crawl of the unt.edu domain by the UNT Libraries for the University Archives. A total of 92,327 PDFs that returned an HTTP response of 200 were present in the archive.  A total of 3,141,886 URIs were present in the entire web archive with PDF content making up just 3\% of the total number of URIs.

A set of 2,000 PDFs were randomly sampled from the full set of PDF documents and were given to two annotators for labeling. These annotators were: one subject matter expert who was responsible for the maintenance of the UNT Scholarly Works Repository and one graduate student with background in Library and Information Systems. They proceeded to label each of the PDF documents as to whether a document would be of interest to the institutional repository or if the document would not be of interest. The labeling of the PDF files resulted in 445 documents (22\%) identified as being of interest for the repository and 1,555 not being of interest. In case of disagreement between annotators, a final decision was made by the researchers of this paper after a discussion with the annotators.

\subsection{Texas.gov dataset}
The next dataset was created from a web archive of websites that constitute the State of Texas web presence.  The data was crawled from 2002 until 2011 and was housed as a collection in the UNT Digital Library. A total of 1,752,366 PDF documents that returned an HTTP response of 200 were present in the archive. A total of 26,305,347 URIs were present in the entire web archive with PDF content making up 6.7\% of the total number of URIs.

As with the first dataset, a random sample of 2,000 PDF documents was given to two annotators for labeling: a subject matter expert from the UNT Libraries and a graduate student (as before). In this case, items were identified as either being in scope for a collection called the ``Texas State Publications Collection'' at the UNT Libraries, or out of scope.  This collection contains a wide range of publications from state agencies.  The labeling of the PDF files resulted in 136 documents (7\%) identified as being of interest for the repository and 1,864 not being of interest. Again, in case of disagreement between annotators, a final decision was made by the researchers of this paper after a discussion with the annotators.

\subsection{USDA.gov dataset}
The last dataset created for this study came from the End of Term (EOT) 2008 web archive. This web archive was created as a collaborative project between a number of institutions at the transition between the second term of George W. Bush and the first term of Barack Obama. The entire EOT web archive contains 160,212,141 URIs. For this dataset we selected the United States Department of Agriculture (USDA) and its primary domain of usda.gov. This usda.gov subset of the EOT archive contains 2,892,923 URIs with 282,203 (9.6\%) of those being PDF files that returned an HTTP 200 response code. 

Similar to the previous datasets, a random sample of 2,000 PDF documents was given to two annotators for labeling: a subject matter expert who has worked as an engineering librarian for a large portion of their career and a graduate student (as before). The subject matter expert has also been involved with the Technical Report Archive and Image Library (TRAIL) that was collecting, cataloging, and digitizing technical reports published by, and for, the federal government throughout the 20th century. The annotators labeled each of the PDF files as either being of interest for inclusion in a collection of Technical Reports or not being of interest to that same collection. The final labeled dataset has 234 documents (12\%) marked as potential technical reports and 1,766 documents identified as not being technical reports. The disagreements between the annotators were finally adjudicated by one of the researchers of this paper. 

The three datasets represent a wide variety of publications that would be considered for inclusion into their target collections. Of these three, the Texas.gov content is the most diverse as the publications range from strategic plans and financial audit reports (which are many pages in lengths) to pamphlets and posters (which are generally very short). The UNT.edu dataset contains publications that are typical for an institutional repository such as research articles, white papers, slide decks from presentations, and other scholarly publications. The publications from the UDSA.gov dataset are similarly scoped as the UNT.edu content, but they also contain a wider variety of content that might be identified as a ``technical report.'' A goal in the creation of the datasets used in this research was to have a true representative sample of the types of content that are held in collections of this kind.

\section{Methods}
\label{sec:methods}
Our goal in this paper is to study different types of features and learning models to accurately distinguish documents of interests from web archives 
for indexing in a specialized collection. In this section, we discuss different types of features that we used in conjunction with traditional machine learning classifiers for finding documents of interests, as well as the Convolutional Neural Network model that do not require any feature engineering.

\subsection{Bag of Words (BoWs)}
``Bag of words'' (BoW) is a simple fixed-length vector representation of any variable length text based on the occurrence of words within the text, with the information about the positions of different words in a document being discarded. First, a vocabulary from the words in the training documents is generated. Then, each document is represented as a vector based on the words in the vocabulary. 
The values in the vector representation are usually calculated as normalized term frequency ({\em tf}) or term frequency - inverse document frequency ({\em tf-idf}) of the corresponding word calculated based on the given document/text.

We experiment with BoW extracted from the full text of the documents as well as from only some portions of documents. Our intuition behind using only some portion of the documents is that many types of documents contain discriminative words at the beginning and/or at the end. For selecting these portions of documents, we consider first-$X$ words from each document, and first-$X$ words combined with last-$X$ words from each document before any type of preprocessing was performed. We experimented with values of $X \in \{100, 300, 500, 700, 1000, 2000\}$. 
For documents with less than $2\cdot X$ words, we considered the entire document without repeating any parts/words from the document. 

Moreover, for BoW encoded from the full text of documents, we also compared the performance of the top-$N$ selected features, 
using the information gain (IG) feature selection method, where $N\in \{300, 500, 1000, 2000, 3000\}$, with the performance of all features. 

\subsection{Structural features}
Structural features (Str) are designed to incorporate aspects specific to documents' structure and are shown to be highly indicative of the classification of academic documents into their document types such as Books, Slides, Theses, Papers, CVs, and Others \shortcite{caragea2016document}. 
%
These features can be grouped into four categories:  file  specific  features,  text  specific features, section specific features, and containment features. Each of these feature categories are described below.

\setlength{\parskip}{0.5em}
\noindent
\textbf{File specific features} include the characteristics of a document such as the number of pages and the file size in kilobytes.

\setlength{\parskip}{0.5em}
\noindent
\textbf{Text specific features} include specifics of the text of a document: the length in characters; the number of words; the number of lines; the average number of words and lines per page; the average number of words per line; the count of reference mentions; the percentage of reference mentions, spaces, uppercase letters, symbols; the ratio of length of shortest to the longest line; the number of lines that start with uppercase letters; the number of lines starting with non-alphanumeric letters; the number of words that appear before the reference section.

\setlength{\parskip}{0.5em}
\noindent
\textbf{Section specific features} include section names and their position within a document. These features are boolean features indicating the appearance of ``abstract'',  ``introduction'',  ``conclusion'',  ``acknowledgements'',  ``references''  and  ``chapter,'' respectively, as well as numeric features indicating position for each of these sections. These features also include two binary features indicating the appearance of ``acknowledgment'' before and after ``introduction.''

\setlength{\parskip}{0.5em}
\noindent
\textbf{Containment features} include containment of specific words or phrases in a document. These features include binary features indicating the appearance of ``this paper,'' ``this book,'' ``this thesis,'' ``this chapter,'' ``this document,'' ``this section,'' ``research interests,'' ``research experience,''  ``education,''  and  ``publications,'' respectively. These features also include three numeric features indicating the position of  ``this  paper,''  ``this  book,''  and  ``this  thesis'' in a document.

Similar to BoW, for the structural features (which are 43 in total), we also compared the performance of the top-$N$ selected features, ranked using the information gain (IG) feature selection method, where $N\in \{10, 20, 30\}$, with the performance of all 43 features. 

\vspace{-2mm}
\subsection{Convolutional Neural Networks (CNNs)}
Convolutional Neural Networks (CNN or ConvNets)~\cite{lecun1998gradient} are a special kind of neural networks to process grid-like structured data, e.g.,  image data. CNNs are associated with the idea of a ``moving filter.'' A convolution consists of a filter or a kernel, that is applied in a sliding window fashion to extract features from the input. This filter is shifted after each operation over the input by an amount called strides. 
The convolution layer consists of multiple filters of different region sizes that generate multiple feature maps for different region sizes. 
Pooling is usually used after the convolution layer to modify the output or reduce the dimensionality. The common practice is to extract the most important feature within each feature map \cite{collobert2011natural,kim2014convolutional}, called 1-max pooling.
Max pooling is applied over each feature map and the maximum values from each filter are selected. Maximum values from each feature map are then concatenated and used as input to a fully connected layer for the classification task. Generally, pooling helps to make the representation become approximately invariant to small changes in the input.
\begin{figure}[h]
    \centering
    \includegraphics[width=0.88\linewidth]{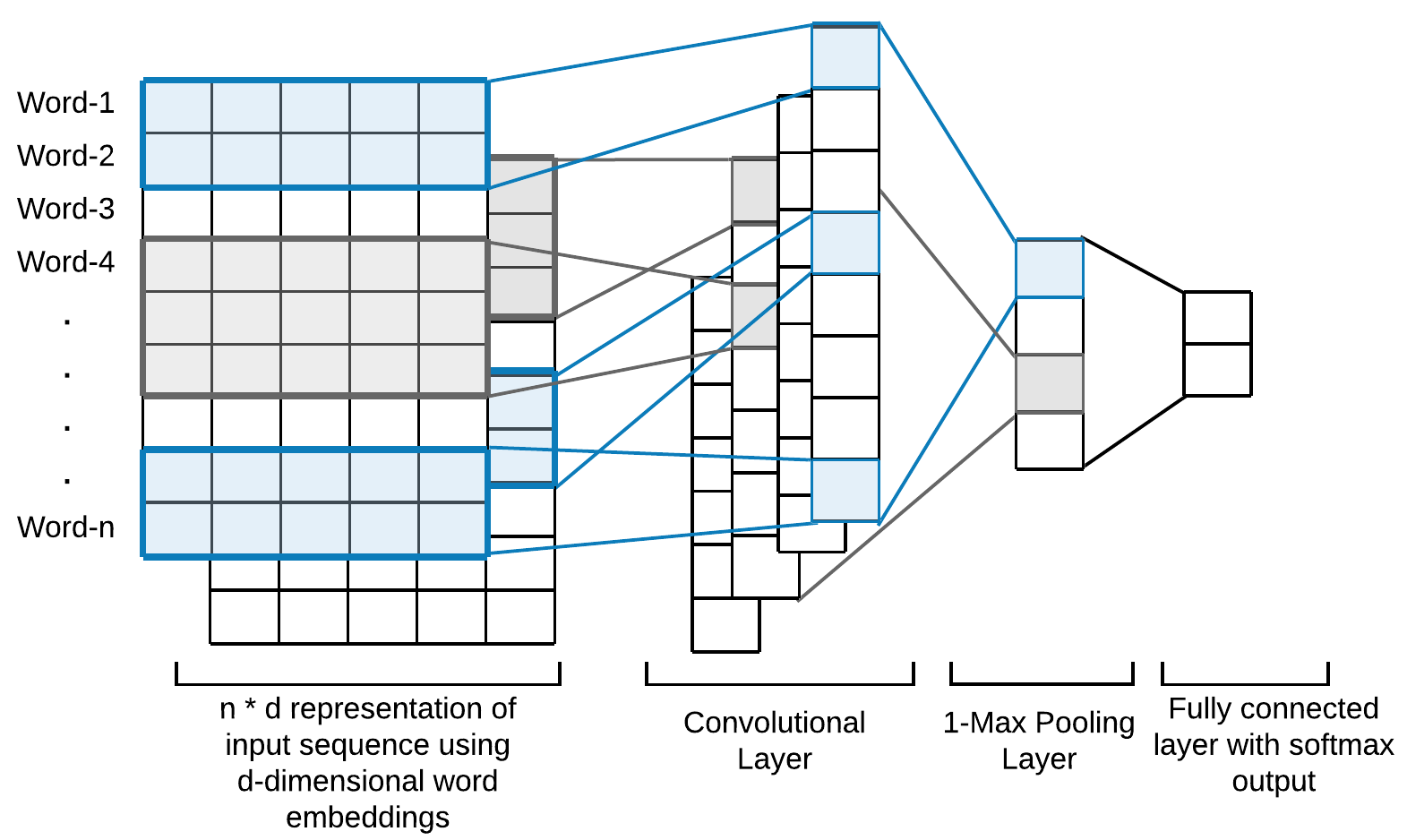}
    \caption{CNN architecture for classification.}
    \label{fig:cnn_arch}
    \vspace{-5mm}
\end{figure}
The CNN architecture that we used in our experiments is shown in Figure~\ref{fig:cnn_arch} and is similar to the CNN architecture developed by Kim \cite{kim2014convolutional}.


For CNN, we experimented with using the text from specific portions of the document. While selecting the portions of documents, as before, we considered first-$X$ words and first-$X$ words combined with last-$X$ words from each document before any preprocessing was performed (where $X \in \{100, 300, 500, 700, 1000\}$). 
For the documents with less than $2\cdot X$ words, we considered the whole document without repeating any part/words from the document.



\section{Experimental Setup and Results}
\label{sec:exp_setup}



In this section, we first discuss the experimental setup of our document classification task (i.e., documents being of interest to a collection or not) and then present the results of our experiments.


\subsection{Experimental setup}

To understand what kind of features are more informative for identifying the  documents of interest for a collection, 
we experiment with the ``bag of words'' (BoW) extracted from the full text of the documents as well as with the text from specific portion of the documents, and with the 43 structural features that capture the structure of the documents. 
For the BoW and structural features (Str) extracted from entire documents (full text), we also compare the performance of top-$N$ 
selected features, which were obtained by using the information gain (IG) feature selection method. For the BoW, we vary $N$ ranging from 300 to 3000, and for the structural features, we experimented with top 10, 20, and 30 selected features by IG. 
%
For the preprocessing step of the BoW, we remove stop words and punctuation, and perform stemming. In addition, we keep only words that appear in at least 5 documents (i.e., having document frequency $df\geq5$). 

\begin{figure*}[!htbp]
\begin{subfigure}{.33\textwidth}
  \includegraphics[width=\linewidth]{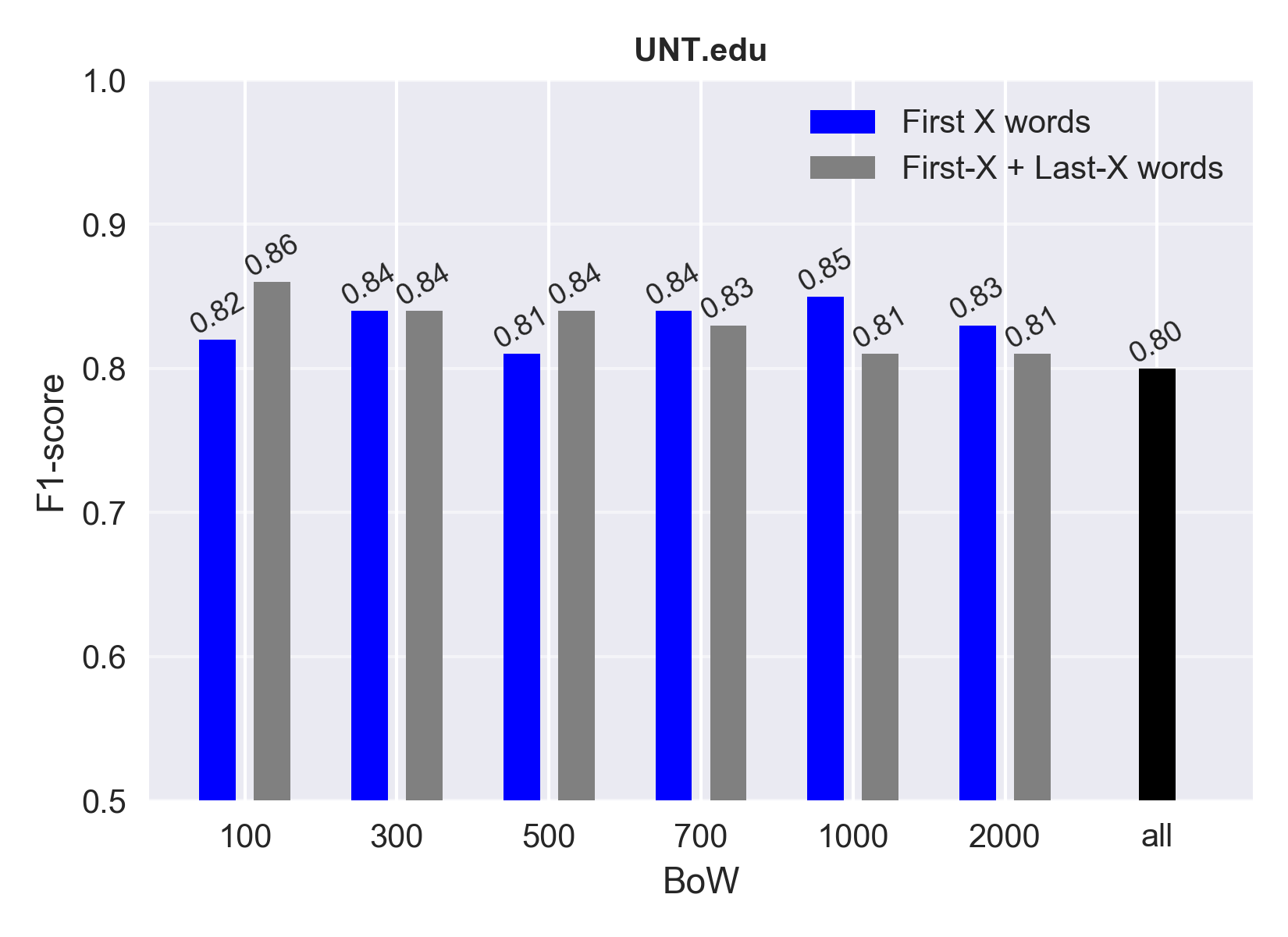}
  \caption{UNT.edu dataset}\label{fig:unt_partdoc_bow}
\end{subfigure}
\begin{subfigure}{.33\textwidth}
  \includegraphics[width=\linewidth]{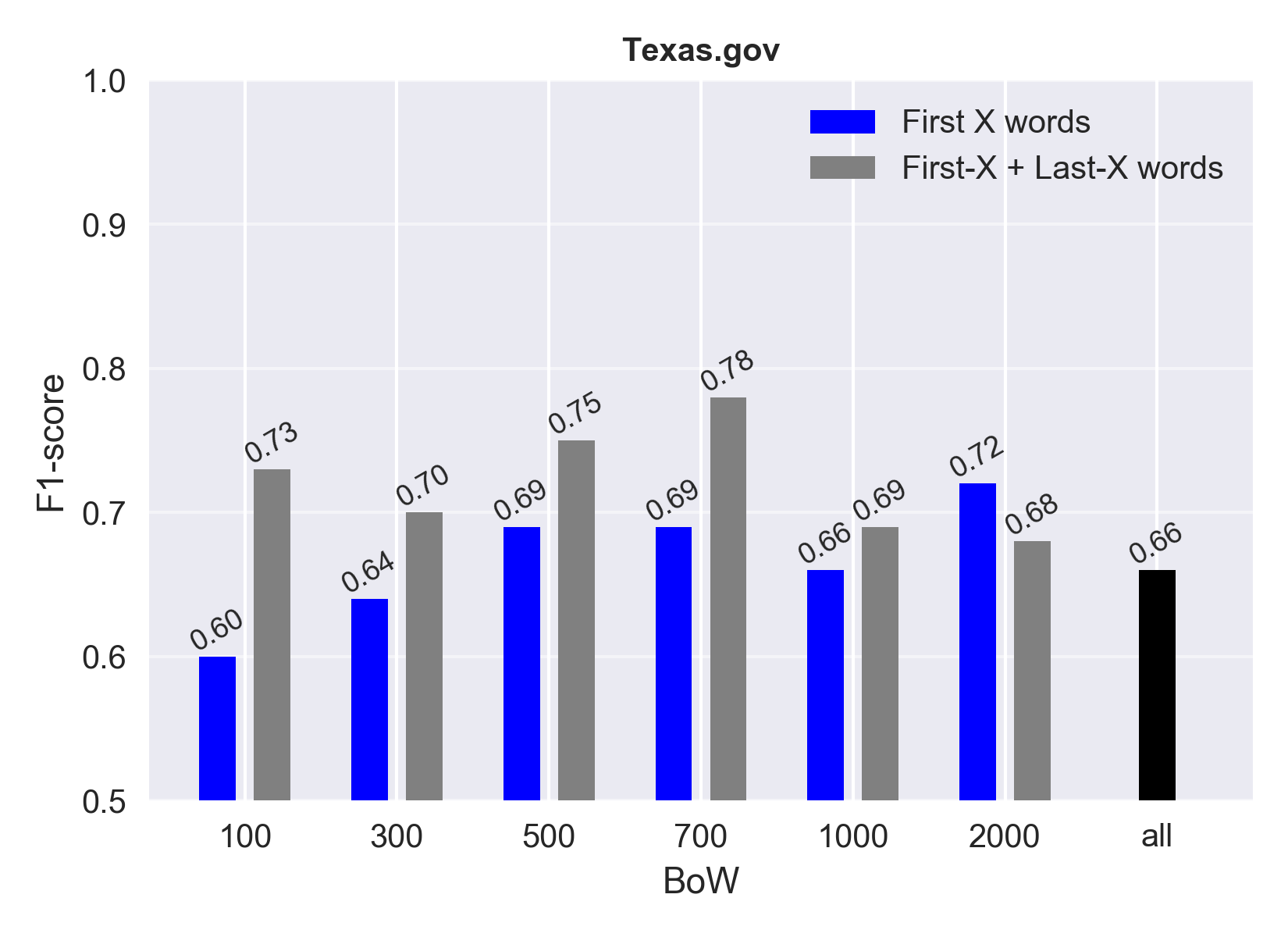}
  \caption{Texas.gov dataset}\label{fig:texas_partdoc_bow}
\end{subfigure}
\begin{subfigure}{.33\textwidth}
  \includegraphics[width=\linewidth]{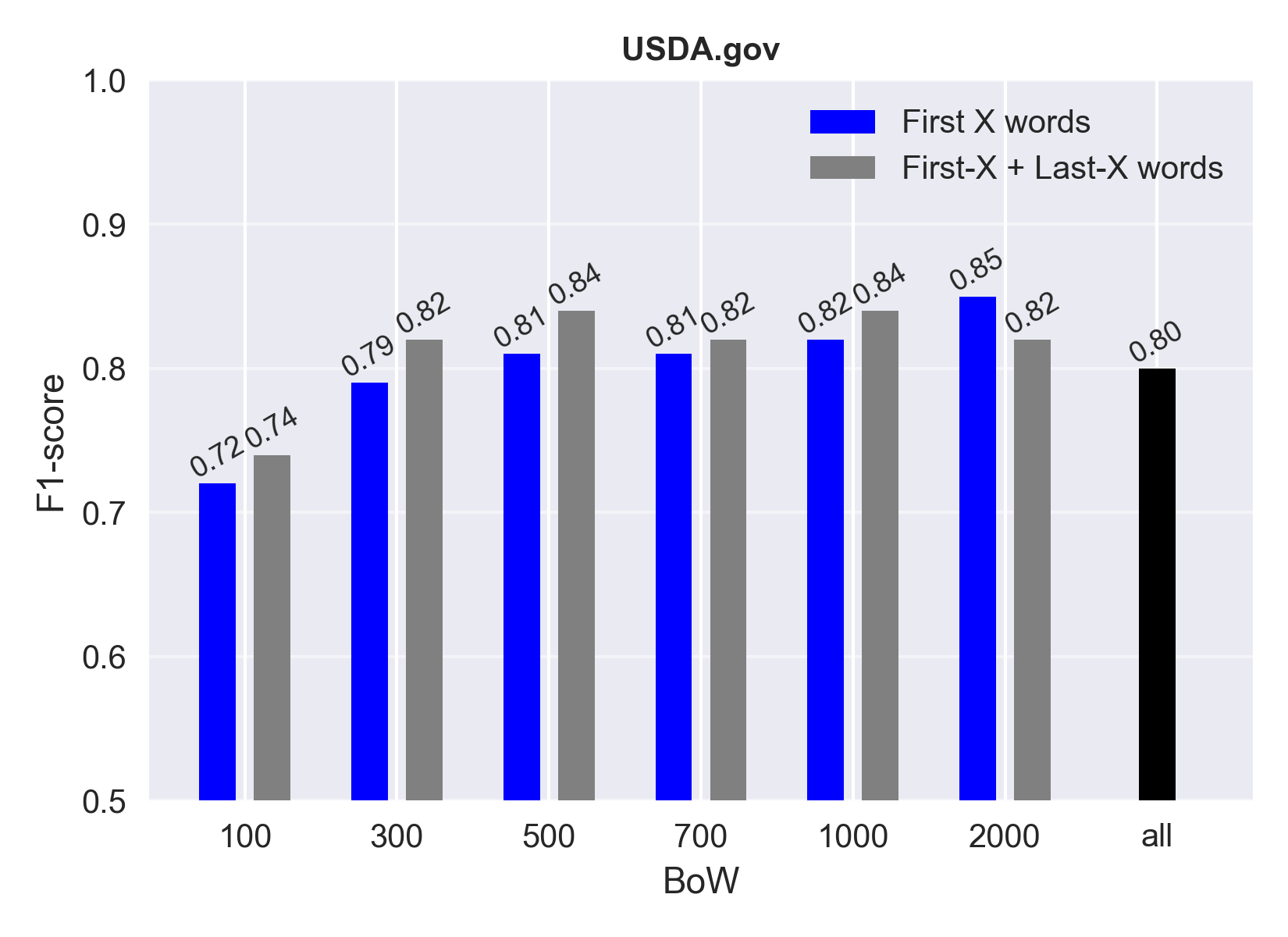}
  \caption{USDA.gov dataset}\label{fig:usda_partdoc_bow}
\end{subfigure}
\vspace{-3mm}
\caption{Performance of BoW classifiers that use various portions of the text of the documents and full text (denoted as `all').} 
\vspace{-3mm}
\label{fig:partdocs_bow}
\end{figure*}

\begin{figure*}[!htbp]
\begin{subfigure}{.33\textwidth}
  \includegraphics[width=\linewidth]{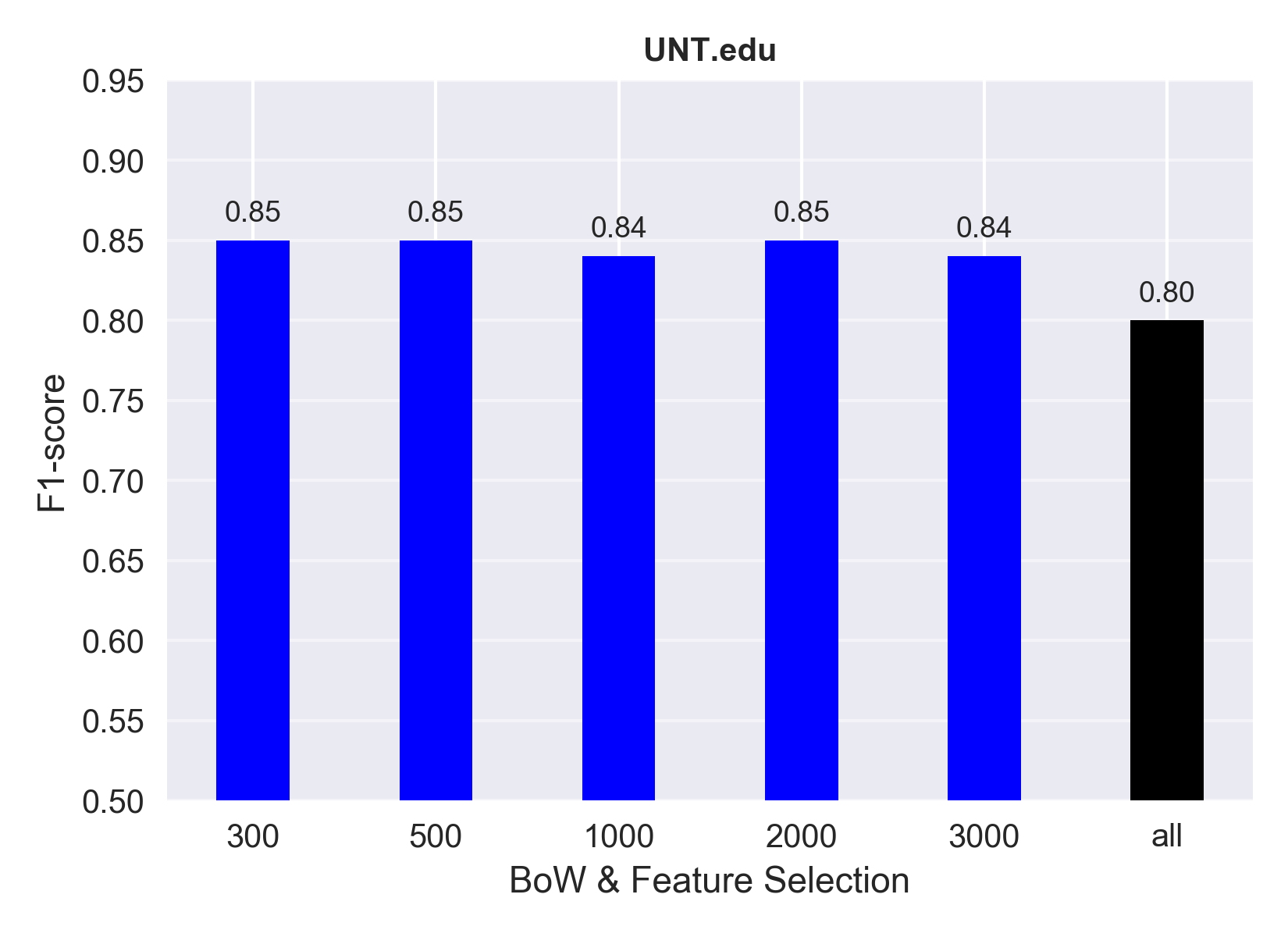}
  \caption{UNT.edu dataset}\label{fig:unt_bow}
\end{subfigure}
\begin{subfigure}{.33\textwidth}
  \includegraphics[width=\linewidth]{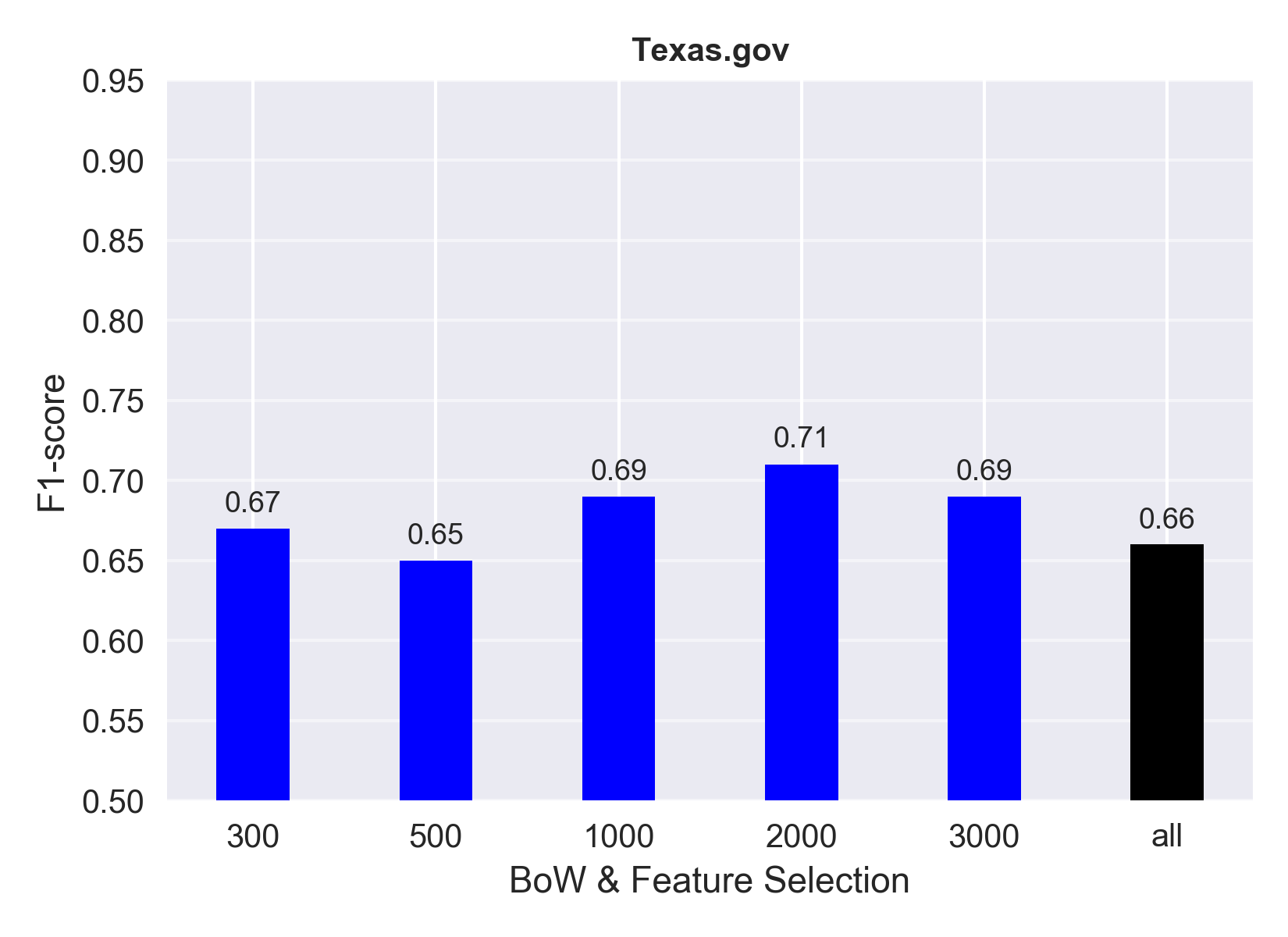}
  \caption{Texas.gov dataset}\label{fig:texas_bow}
\end{subfigure}
\begin{subfigure}{.33\textwidth}
  \includegraphics[width=\linewidth]{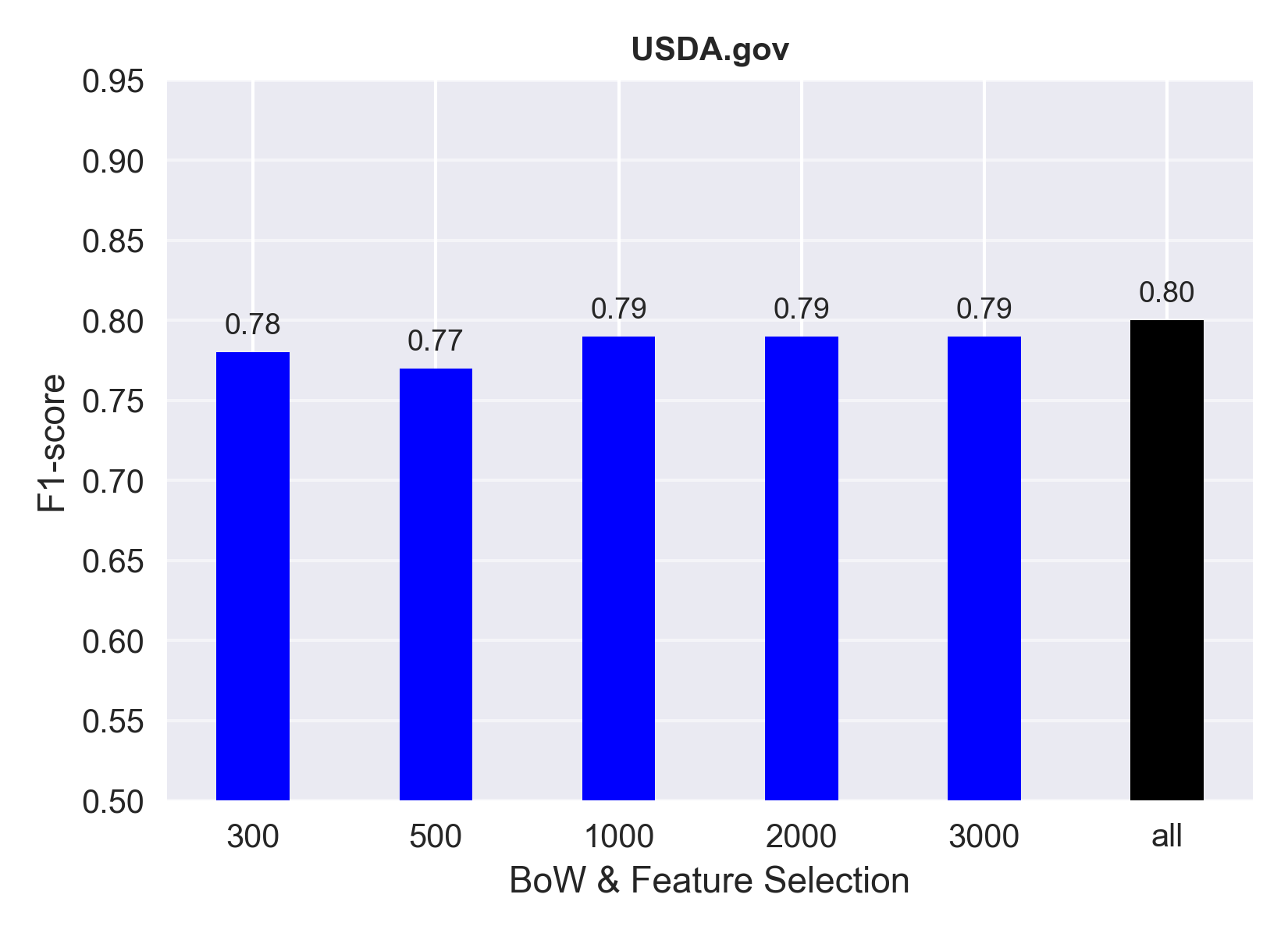}
  \caption{USDA.gov dataset}\label{fig:usda_bow}
\end{subfigure}
\vspace{-3mm}
\caption{Performance of BoW classifiers that use the full text of the documents (`all') and its feature selection.}
\vspace{-3mm}
\label{fig:bow}
\end{figure*}

Using the above features, we experiment with several traditional 
machine learning classifiers:  Gaussian Naive Bayes (GNB)~\cite{andrew:nb}, Multinomial Naive Bayes (MNB)~\cite{andrew:nb}, Random  Forest  (RF) \cite{mlj01breiman},  Decision  Trees  (DT)~\cite{quinlan1986induction}, and Support Vector Machines with a linear kernel (SVM)~\cite{Joachims1998}. We used the scikit-learn\footnote{https://scikit-learn.org/stable/} implementation of these classifiers.

In addition to these models, we also investigate the performance of Convolutional Neural Networks (CNNs) on our task. 
Our CNNs comprise of mainly two layers (as shown in Figure \ref{fig:cnn_arch}): a convolutional layer followed by max pooling and a fully connected layer for the classification. For the CNN input, we consider a document (partial) as a sequence of words and use pre-trained word embeddings for each word. These pre-trained word embeddings are trained on the Google News dataset using the Word2Vec\footnote{ \url{https://code.google.com/archive/p/word2vec/}}~\cite{mikolov2013distributed} algorithm. 
For CNN, we used its TensorFlow implementation.\footnote{\url{https://www.tensorflow.org/}}



\setlength{\parskip}{0.5em}
\noindent
\textbf{Train, Development, Test Splits.} 
From the original datasets of 2,000 PDF files, we divided each dataset into three parts by randomly sampling training set (\textbf{Train}), development set (\textbf{Dev}), and test set (\textbf{Test}) 
from each dataset. All \textbf{Train}, \textbf{Dev}, and \textbf{Test} follow a similar distribution as the original dataset.
Table~\ref{tab:ds} shows the number of positive (+) and negative (-) examples (i.e., documents of interest or not of interest, respectively) in each of the three datasets 
for which we were able to extract the text (from a given PDF document). 
For our purpose, to extract the text from the PDF documents, we used PDFBox.\footnote{http://pdfbox.apache.org/} The scanned documents and other documents for which the text was not correctly extracted were ignored.

\begin{table}[!h]
	\begin{center}
				\begin{tabular}{|l||r|r||r|r||r|r|}
					\hline
                 \multirow{2}{*}{} & \multicolumn{2}{r||} {UNT.edu}  & \multicolumn{2}{c||} {Texas.gov}  & \multicolumn{2}{r|} {USDA.gov} \\
                \cline{2-7}
               Datasets & $-$ & + & $-$ & + & $-$ & + \\
					\hline
                    \hline
    \hline
	Train & 869 & 250 & 981 & 72 & 907 & 121 \\
    {\bf Dev}  & 290 & 83 & 327 & 24 & 300 & 40\\
    {\bf Test}  & 290 & 83 & 327 & 24 & 300 & 40\\
	\hline
	\textbf{Train-2} & 869 & 434 & 981 & 490 & 907 & 453 \\
    \hline
				\end{tabular}
		\end{center}
		\caption{Datasets description.}
        \vspace{-18pt}
		\label{tab:ds}
\end{table}

Because the original datasets are very skewed (see Section 3), with only around 22\%, 7\%, and 12\% of the PDF documents being part of the positive class (i.e., to be included in a collection), we asked the subject matter experts of each web archive collection to further identify more positive examples. The supplemental positive examples (for each collection) were added to the training set of the corresponding collection. 
Specifically, for the training set, we sampled from the newly labeled set of positive examples so that the number of negative examples is doubled as compared with the number of positive examples. We denote this set as \textbf{Train-2} (see Table~\ref{tab:ds}). 
Note that the test and dev sets of each collection remained the same (i.e., having the original distribution of the data to mimic a real world scenario in which data at test come is fairly skewed). 

Note that we studied the performance of our models using other positive to negative data distributions in the training set. 
%
However, we found that the models trained on \textbf{Train-2} perform better than when we train on other data distributions (e.g., the original or 1:1 data distributions). We estimated this  
on the development set that we constructed as explained above. Thus, in the next sections, we report the results when we train on \textbf{Train-2} (2:1 distribution) and evaluate on {\bf Test} (original distribution). 

Moreover, we use the development set also for the hyper-parameter tuning for the classifiers, and for the best classifier selection (which in our experiments was a Random Forest).
In experiments, we tuned hyper-parameters for different classifiers as follows: the C parameter in SVM $\in \{0.01, 0.05, 0.1\}$; the number of trees in RF $\in \{20, 23, 25, 27, 30\}$; in CNN, single as well as three types of filters with region sizes $\in \{1, 3, 4, 5, 7
\}$; the number of each type of filters in CNN $\in \{100, 128, 200\}$. 

\textbf{Evaluation Measures.}
To evaluate the performance of various classifiers, we use precision, recall, and F1-score for the positive class. 
All experiments are repeated three times with a different train/dev/test split obtained using three different random seeds, and the final results are averaged across the three runs. 
We first discuss the results in terms of the F1-score  using bar plots. Then we present all measures: precision, recall, and F1-score, in a table. 



\begin{table*}
\parbox{.30\linewidth}{
\centering
                \begin{tabular}{|l|c|c|c|}
				\hline
				 \multicolumn{4}{|c|} {\textbf{BoW}} \\
				\hline
                & UNT.edu & Texas.gov & USDA.gov \\
				\hline
	            1 & data & www & studi \\
                2 & al & texa & method \\
                3 & result & program & research \\
                4 & figur & tx & result \\
                5 & compar & area & al \\
                6 & increas & ag & effect \\
                7 & similar & includ & potenti \\
                8 & rang & year & observ \\
                9 & semest & inform & occur \\
                10 & larg & system & found \\
                11 & tabl & site & measur \\
                12 & model & public & speci \\
                13 & conclusion & nation & water \\
                14 & research & contact & larg \\
                15 & measur & result & determin \\
                16 & recent & import & similar \\
                17 & abstract & number & environ \\
                18 & exist & manag & high \\
                19 & show & reduc & natur \\
                20 & low & increas & introduc \\
                21 & comparison & continu & differ \\
                22 & de & level & increas \\
                23 & high & servic & reduc \\
                24 & usa & plan & analysi \\
                25 & observ & base & environment \\
                26 & doi & qualiti & signific \\
                27 & base & state & suggest \\
                28 & signific & work & experi \\
                29 & lack & time & control \\
                30 & suggest & design & site \\
                \hline
				\end{tabular}
		\caption{Top-30 selected features from the BoW using information gain.}
        \vspace{-3mm}
		\label{tab:top_bow_ig}
}
\hfill
\parbox{.60\linewidth}{
\centering
                \begin{tabular}{|l|c|c|c|}
				\hline
				 \multicolumn{4}{|c|} {\textbf{Structural Features}} \\
				\hline
                & UNT.edu & Texas.gov & USDA.gov \\
				\hline
	            1 & positionOfThisPaper & fileSize & refCount \\
                2 & refCount & numLines & refRatio \\
                3 & refRatio & lnratio & strLength \\
                4 & positionOfReferences & strLength & positionOfThisPaper \\
                5 & fileSize & pgCount & pgCount \\
                6 & tokBeforeRef & numTok & numTok \\
                7 & references & thisDocument & positionOfIntro \\
                8 & pgCount & publications & intro \\
                9 & positionOfAbstract & positionOfIntro & numLines \\
                10 & thisPaper & intro & positionOfAbstract \\
                11 & concl & education & positionOfReferences \\
                12 & positionOfConcl & positionOfThisPaper & references \\
                13 & strLength & avgNumLinesPerPage & abstract \\
                14 & numLines & symbolStart & tokBeforeRef \\
                15 & abstract & ucaseStart & ucaseStart \\
                16 & numTok & spcRatio & fileSize \\
                17 & positionOfIntro & avgNumWordsPerLine & positionOfConcl \\
                18 & ucaseStart & refRatio & concl \\
                19 & positionOfAck & positionOfAck & spcRatio \\
                20 & ack & ack & positionOfAck \\
                21 & avgNumWordsPerPage & positionOfConcl & ack \\
                22 & avgNumLinesPerPage & concl & symbolRatio \\
                23 & AckAfterIntro & positionOfReferences & ucaseRatio \\
                24 & symbolRatio & positionOfAbstract & thisPaper \\
                25 & spcRatio & tokBeforeRef & symbolStart \\
                26 & symbolStart & references & AckAfterIntro \\
                27 & lnratio & refCount & lnratio \\
                28 & ucaseRatio & avgNumWordsPerPage & avgNumWordsPerPage \\
                29 & intro & ucaseRatio & avgNumWordsPerLine \\
                30 & publications & AckBeforeIntro & avgNumLinesPerPage \\
                \hline
				\end{tabular}
		\caption{Top-30 selected features from the 43 structural features using information gain.}
        \vspace{-3mm}
		\label{tab:top_43_ig}
}
\end{table*}

\subsection{The performance of BoW and its feature selection}

{\bf BoW Performance.} First, we compare the performance of the BoW classifiers when we use various portions of the text of the documents with that of the BoW classifiers that use the full text of the documents. For the various portions of the text, we use first $X$ words and first-$X$ words combined with last-$X$ words from each document, where $X \in \{100, 300, 500, 700, 1000, 2000\}$. 
%
The results of this set of experiments are shown in Figure \ref{fig:partdocs_bow} for all three datasets, UNT.edu, Texas.gov, and USDA.gov, respectively. 
Random Forest performs best among all classifiers for the BoW features, and hence, we show the results using only Random Forest. 

As can be seen from Figure~\ref{fig:unt_partdoc_bow}, on UNT.edu, the BoW that uses the first-100 words combined with last-100 words from each document performs best compared with the performance of the BoW that uses other parts of the documents and achieves a highest F1-score of $0.86$. 
Interestingly, the BoW that uses the entire text of documents performs worse than the BoW that focuses only on specific portions of the text of each document, i.e., the BoW that uses the entire text of the documents achieves a lowest F1-score of $0.80$ as compared with $0.86$ achieved by BoW that uses only the first-100 + last-100 words from each document. This means that using the entire text of a document introduces redundant or irrelevant features that are not beneficial for the classification task. 

On the Texas.gov dataset, it can be seen from  Figure~\ref{fig:texas_partdoc_bow} that the performance of the BoW classifiers increases as we add more words from the beginning and the end of each document up to 700 words and after that the performance starts decreasing. The BoW classifier that uses the first-700 words combined with last-700 words from each document achieves a highest F1-score of $0.78$. On the other hand, the BoW that uses the entire text of documents mostly performs worse than the BoW that focuses only on specific portions of the documents, e.g., the BoW that uses the entire content of the documents achieves an F1-score of $0.66$ as compared with the BoW that uses only the first-700 + last-700 words, which achieves an F1-score of $0.78$. On Texas.gov, the BoW classifiers that use words from the beginning and ending of the documents generally outperform those that use words only from the beginning of the documents.

On USDA.gov, as can be seen from  Figure~\ref{fig:usda_partdoc_bow}, the BoW classifiers that use words from the beginning combined with the ending of the documents (first-$X$ + last-$X$ words) generally outperform the BoW classifiers that use words only from the beginning of documents. These results are similar to those obtained on Texas.gov, although the difference in performance is much smaller compared with Texas.gov.
However, interestingly, we notice that the BoW classifier that uses only the first-2000 words performs best and achieves an F1-score of $0.85$, which is followed closely by the BoW classifier . As before, the BoW that uses the entire text of documents usually performs worse than the BoW that focuses only on specific portions of each document, e.g., the BoW that uses the entire text of the documents achieves an F1 of $0.80$ as compared to $0.85$ achieved by BoW on first-2000 words from the documents. 

From Figure \ref{fig:partdocs_bow}, we can also notice that the performance of BoW classifiers on Texas.gov is lower compared with that of classifiers on UNT.edu and USDA.gov, which could be explained by a higher diversity in Texas.gov compared with the other two collections.

\begin{figure*}[!htbp]
\begin{subfigure}{.33\textwidth}
  \includegraphics[width=\linewidth]{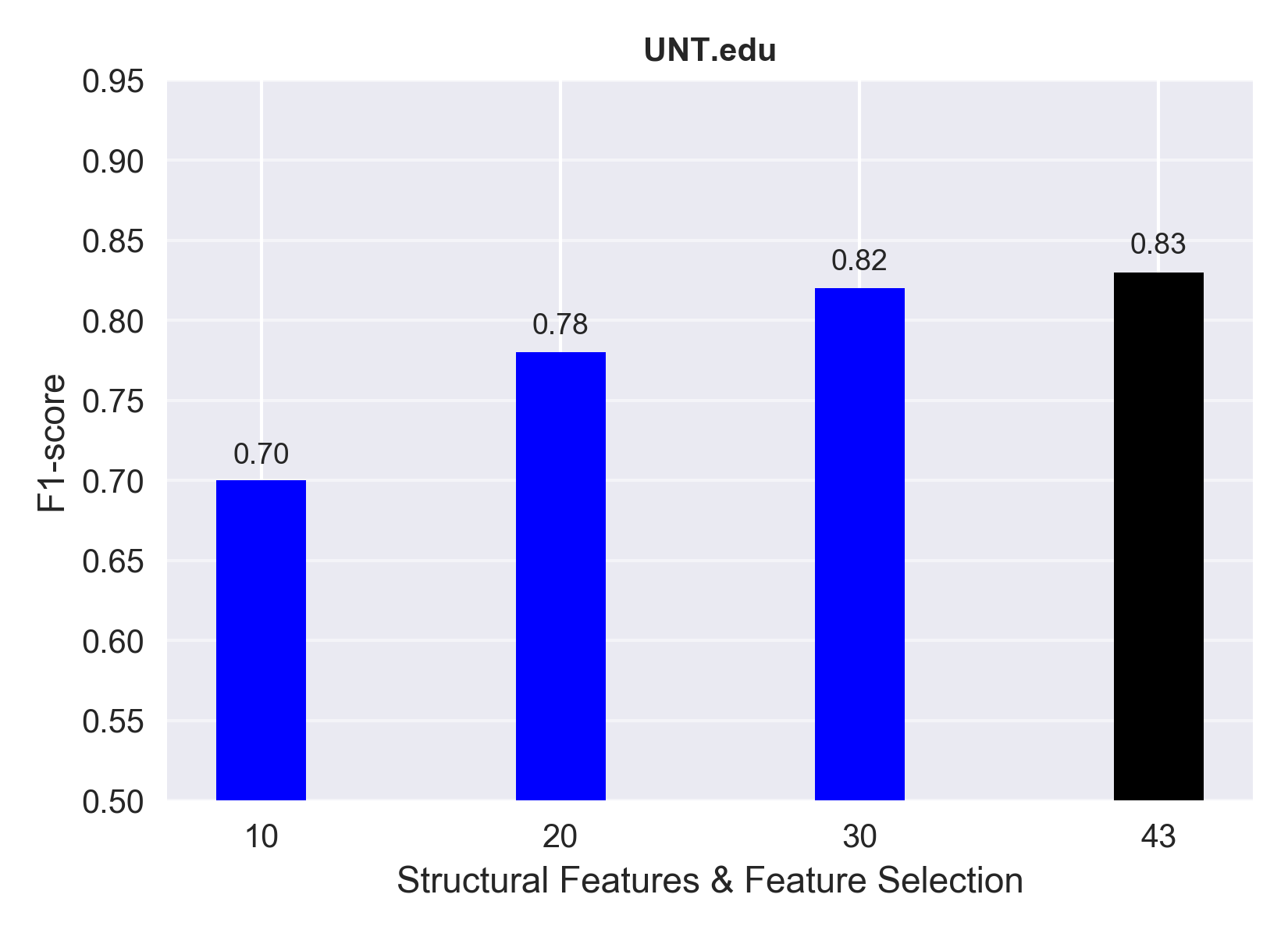}
  \caption{UNT.edu dataset}\label{fig:unt_43}
\end{subfigure}
\begin{subfigure}{.33\textwidth}
  \includegraphics[width=\linewidth]{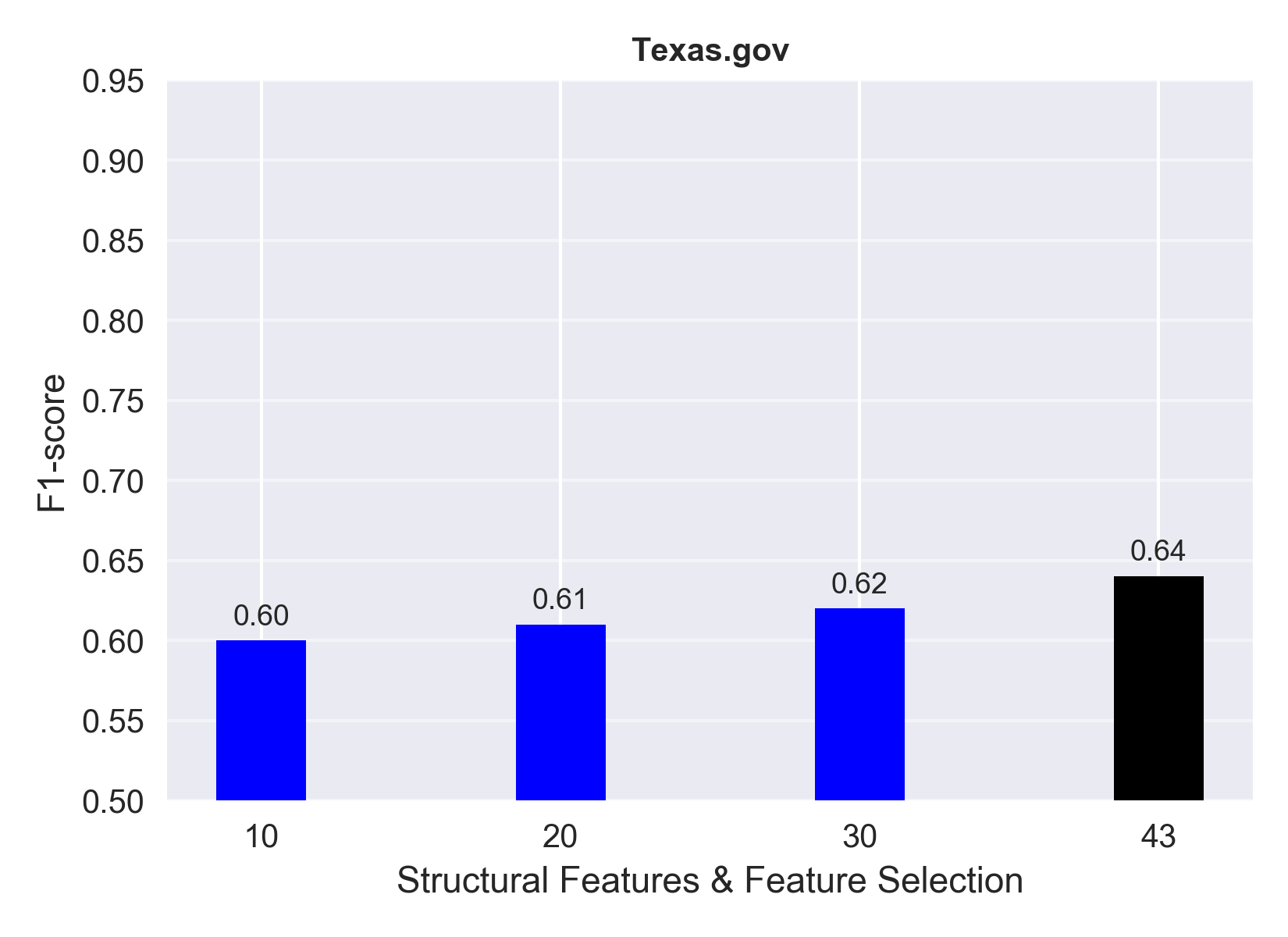}
  \caption{Texas.gov dataset}\label{fig:texas_43}
\end{subfigure}
\begin{subfigure}{.33\textwidth}
  \includegraphics[width=\linewidth]{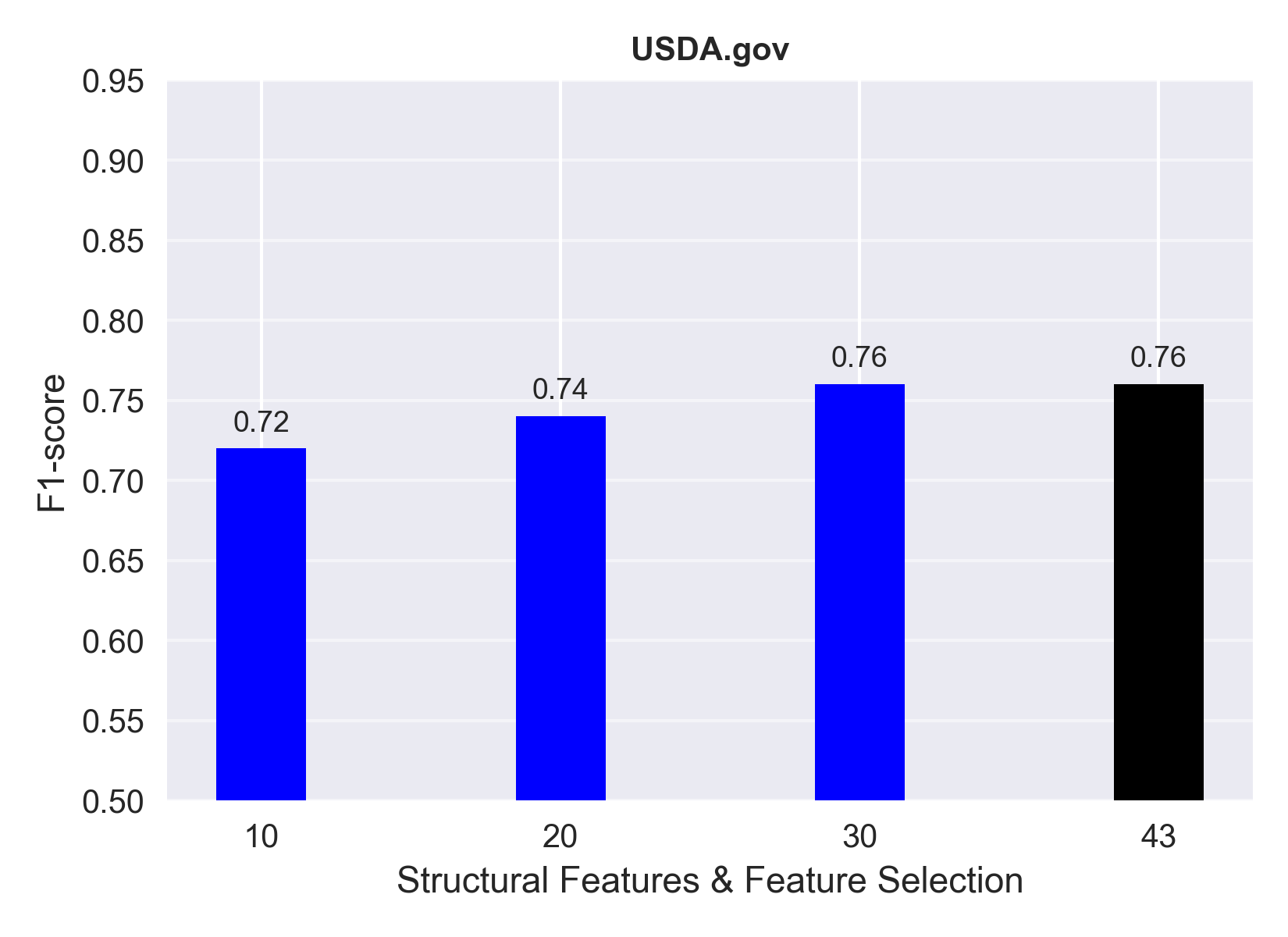}
  \caption{USDA.gov dataset}\label{fig:usda_43}
\end{subfigure}
\caption{Performance of structural features and their feature selection.}
\label{fig:43str}
\vspace{-3mm}
\end{figure*}

\textbf{Feature selection on the BoW extracted from the entire document text.}
Next, we show the effect of feature selection on the performance of BoW classifiers that use the full text of the documents. To rank the features and select the top $N$ best features, we use information gain.  
%
Figure~\ref{fig:bow} compares the performance of the BoW classifiers that use all features (denoted BoW-all) with that of classifiers that use the top-$N$ selected features by information gain, for all three datasets, where $N \in \{300, 500, 1000, 2000, 3000\}$. In total, BoW-all has $19733$, $19625$, and $22255$ features for UNT.edu, Texas.gov, and USDA.gov, respectively.
From the figure, we notice that performing feature selection improves the performance of BoW-all extracted from the full text of documents for UNT.edu and Texas.gov, whereas the performance decreases slightly on USDA.gov. For example, on UNT.edu, top-300 selected features achieve an F1-score of $0.85$ as compared with $0.80$ achieved by BoW-all. 
On Texas.gov, the highest performance is obtained using top-2000 selected features, which achieve a highest F1-score of $0.71$ as compared with $0.66$ achieved by BoW-all.
On USDA.gov, the top-1000, 2000, and 3000 features achieve higher performance as compared to top-300 and top-500 features. Unlike the other two datasets, on the USDA.gov dataset, BoW-all achieves a highest F1 of $0.80$ as compared with other top-$N$ selected features.
Comparing Figure \ref{fig:partdocs_bow} with Figure \ref{fig:bow}, it is interesting to note that, although feature selection improves the performance of BoW-all for UNT.edu and Texas.gov, still the performance of feature selection performed on words from the entire content of documents is not as good as the performance of BoW that uses words from the beginning or the beginning and ending of documents.

Table~\ref{tab:top_bow_ig} shows the top-30 ranked words using information gain feature selection method on each dataset. For the UNT.edu data, we see tokens that appear to be associated with academic publications such as ``data, result, figure, research, or conclusion,'' which seem to match the general scope of this collection as it contains research articles and publications authored by faculty members. The Texas.gov BoW features include tokens that align with publications or other official documents including ``texa (texas), program, area, site, nation, or system.'' These also seem to align very well with the kind of publications selected as being in scope in the dataset. Finally, the USDA.gov BoW selected features include tokens from research and technical publications with tokens such as ``study, method, research, result, and effect.'' There is more overlap between these tokens in USDA.gov and the tokens from the UNT.edu dataset. This suggests that there is some overlap in the kind of content between the two datasets (confirmed by subject matter experts as well). 

Next, we explore the following question: {\textit{Where are the best performing selected features  located in the documents?}} To answer this question, we check the overlap between the best performing top-$N$ selected features and the best performing BoW that uses only specific portions of the text of the documents. For all three datasets, we found that all best performing top-$N$ selected features are present in the best performing BoW that uses only specific portions of the document, e.g., on Texas.gov, all top-2000 selected features occur in the BoW that uses the first-700 + last-700 words from each of the documents.

\subsection{The performance of structural features and their feature selection}
Here, we compare the performance of the 43 structural features with the performance of different top-$N$ selected structural features by information gain. Again, Random Forest performs best compared with any other classifier we experimented with for the structural features and their feature selection. 
Figure~\ref{fig:43str} shows the performance of the 43 structural features (Str) and the top-$N$ selected features by information gain, for all three datasets. As can be seen from the figure, 
for UNT.edu and Texas.gov, the performance of the Str classifiers keeps increasing from top-10 features to all 43 features. As expected, on Texas.gov, the performance of Str classifiers is much lower compared with that of Str classifiers on UNT.edu. This is because the Str features are more aligned with academic documents, whereas Texas.gov covers a more diverse set of documents. 
On USDA.gov, the performance of the Str classifiers keeps increasing from top-10 selected features to top-30 features, and the classifiers corresponding to top-30 and all 43 features perform the same.


Table~\ref{tab:top_43_ig} shows the top-30 ranked structural features using the information gain feature selection method on each dataset. Interpreting these structural features is similar to the BoW results discussed above. The UNT.edu shows that the most informative features include those that are closely aligned with the scholarly publications including positionOfThisPaper, refCount, refRatio, and positionOfReferences. 
The USDA.gov has similar structural features that key off of the content of the publications but also start to include more generic features such as strLength, pgCount, and numTok into the most informative features. The Texas.gov is very different, with the most informative structural features being those that are very generic such as fileSize, numLines, lnratio, strLength, and pgCount.  This seems to match the content in the datasets where UNT.edu is well focused on scholarly publications, USDA.gov includes both scholarly publications as well as technical reports, and Texas.gov is very broad in the kind of publications included in the collection. Because of this broad range of publications in Texas.gov, it appears that the 43 structural features selected are not being used to their fullest capability for this dataset.

\begin{figure*}[!htbp]
\begin{subfigure}{.33\textwidth}
  \includegraphics[width=\linewidth]{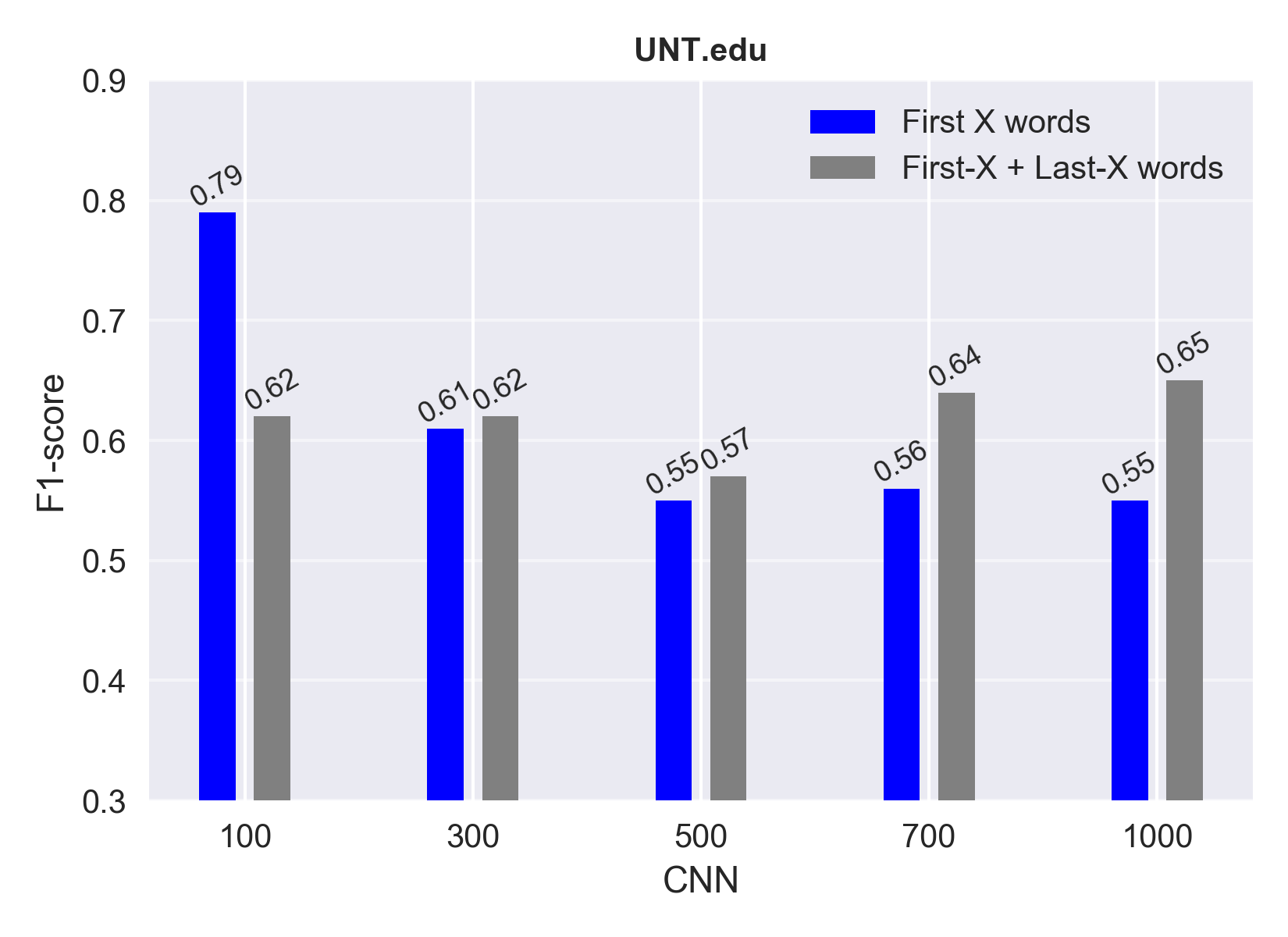}
  \caption{UNT.edu dataset}\label{fig:unt_partdoc_cnn}
\end{subfigure}
\begin{subfigure}{.33\textwidth}
  \includegraphics[width=\linewidth]{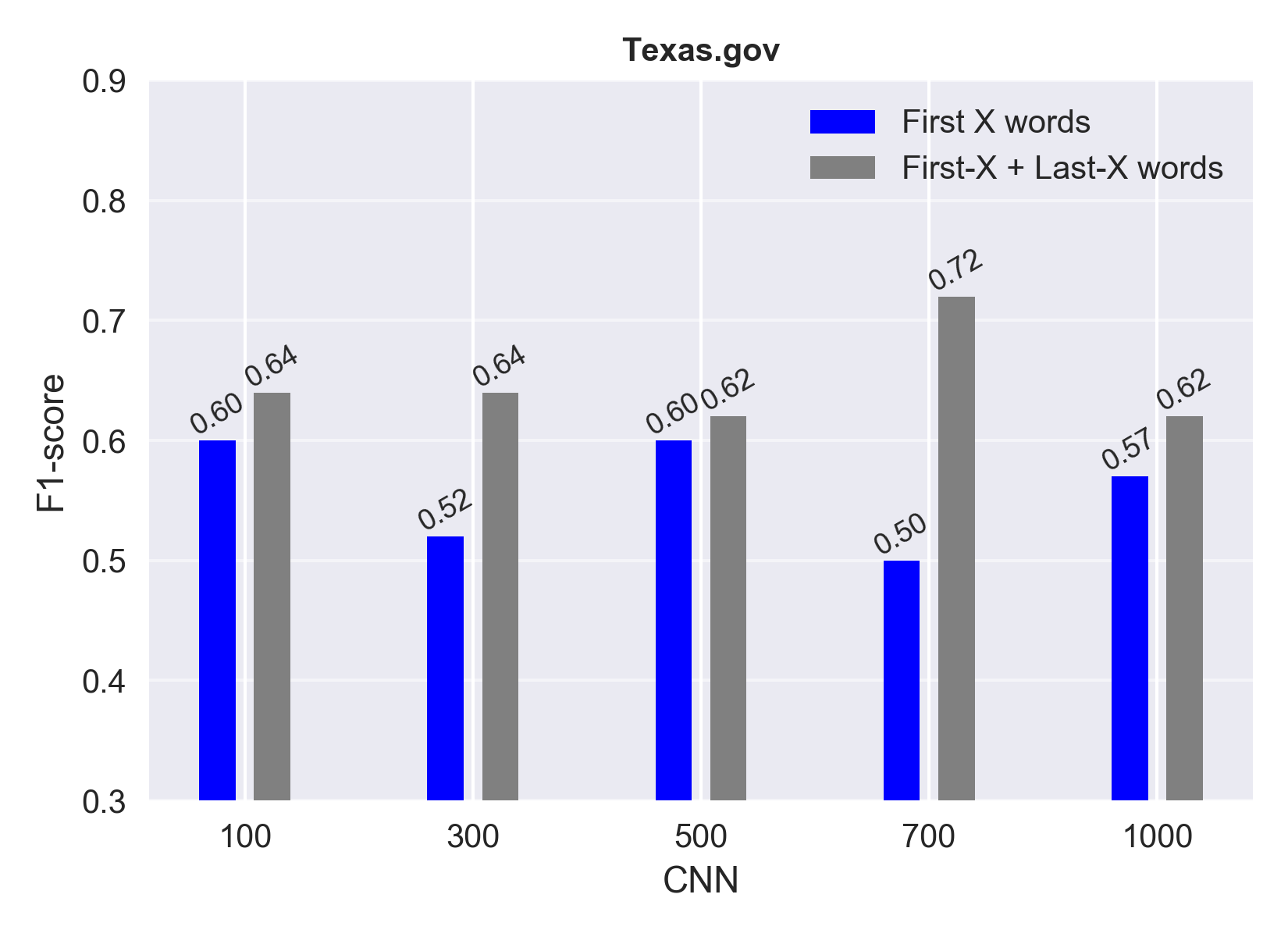}
  \caption{Texas.gov dataset}\label{fig:texas_partdoc_cnn}
\end{subfigure}
\begin{subfigure}{.33\textwidth}
  \includegraphics[width=\linewidth]{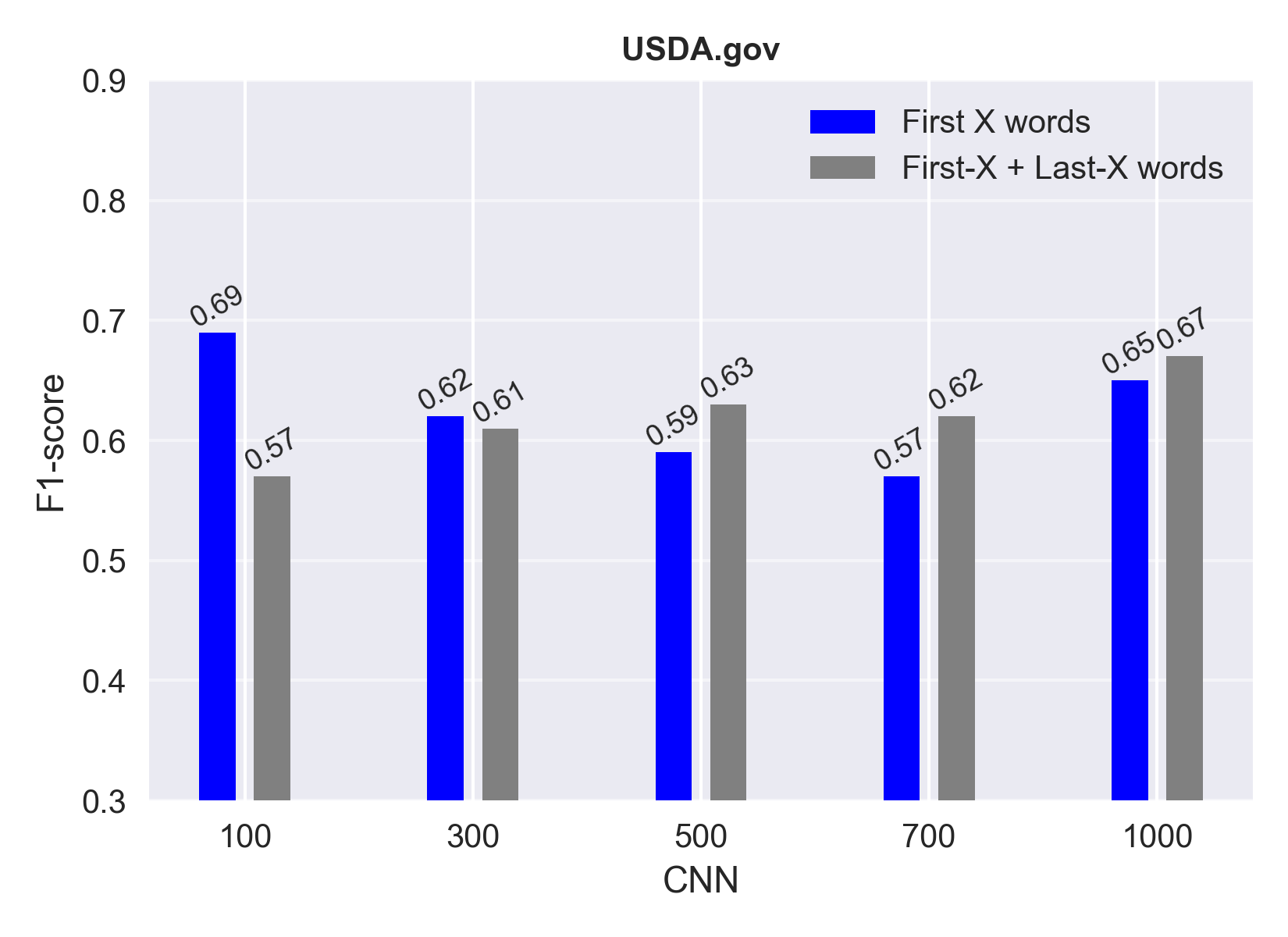}
  \caption{USDA.gov dataset}\label{fig:usda_partdoc_cnn}
\end{subfigure}
\caption{Performance of the CNN using different portions of the documents on different datasets} 
\label{fig:partdocs_cnn}
\end{figure*}

\begin{table*}[!htbp]
	\begin{center}
				\begin{tabular}{|l||c|c|c||c|c|c||c|c|c|}
					\hline
				\multirow{2}{*}{\textbf{Feature Type}} & \multicolumn{3}{c||} {UNT.edu}  & \multicolumn{3}{c||} {Texas.gov}  & \multicolumn{3}{c|} {USDA.gov} \\
                \cline{2-10}
                & Pr & Re & F1 & Pr & Re & F1 & Pr & Re & F1 \\
					\hline
                    \hline
	BoW-all & 0.88 & 0.74 & 0.80 & 0.54 & 0.86 & 0.66 & 0.76 & 0.86 & 0.80 \\
	BoW-PD & \textbf{0.90} & 0.83 & \textbf{0.86} & 0.67 & \textbf{0.91} & \textbf{0.78} & \textbf{0.82} & \textbf{0.88} & \textbf{0.85} \\
	FS-BoW & 0.86 & \textbf{0.84} & 0.85 & 0.61 & 0.86 & 0.71 & 0.71 & \textbf{0.88} & 0.79 \\	
    \hline
	Str	& 0.84 & 0.82 & 0.83 & 0.49 & 0.90 & 0.64 & 0.70 & 0.85 & 0.76 \\
    \hline
    CNN & 0.78 & 0.80 & 0.79 & \textbf{0.89} & 0.61 & 0.72 & 0.65 & 0.75 & 0.69 \\
    \hline
				\end{tabular}
		\end{center}
		\caption{Performance of different features/models on our datasets.}
		\label{tab:results}
\end{table*}

\subsection{The performance of the CNN classifiers}

Next, we compare the performance of the CNN classifiers when we consider the text from different portions of the documents.

Figure~\ref{fig:partdocs_cnn} shows the performance of the CNN classifiers that use word sequences from various portions of the documents, i.e., first $X$ words and first-$X$ words combined with last-$X$ words (where $X \in \{100, 300, 500, 700, 1000\}$).  

On UNT.edu, it can be seen from  Figure~\ref{fig:unt_partdoc_cnn} that the CNN that uses the first-100 words from each of the documents performs best and achieves a highest F1-score of $0.79$. Also, on UNT.edu, the CNNs that use words from the beginning and ending of each document generally outperform the CNNs that use words only from the beginning of the documents (except for 100-length sequences). Moreover, we notice the drastic performance gap between the performance of the CNN that uses the first-100 words and the CNNs that use other portions of the documents' text.

On Texas.gov, it can be seen from  Figure~\ref{fig:texas_partdoc_cnn} that the CNN classifier that uses the first-700 words combined with the last-700 words performs best and achieves an F1-score of $0.72$. For Texas.gov, the CNN classifiers that use words from the beginning and ending portions of documents outperform the CNNs that use words only from the beginning of documents. 

On USDA.gov, we can see from Figure~\ref{fig:usda_partdoc_cnn} that the CNN classifier that uses the first-100 words from each document performs best and achieves a highest F1-score of $0.69$. The performance of the CNN classifiers that use word sequences from the beginning and ending of documents perform better than those that use only the beginning of documents for word sequence lengths greater than 500.
Comparing the results in Figure \ref{fig:partdocs_cnn} with the previous results, we can notice that the deep learning CNN models perform worse than the Random Forest classifiers that use BoW from various portion of the document text.

\vspace{-3mm}
\subsection{Overall comparison}

Last, we contrast the performance of different sets of features and models in terms of all compared measures, precision (Pr), recall (Re), and F1-score (F1) for the positive class. 
Table~\ref{tab:results} shows the BoW classifier that uses the full text of each document (\textbf{BoW-all}), the BoW classifier that uses some portion of the document (\textbf{BoW-PD}), the best performing top-$N$ selected features from the BoW-all using information gain (\textbf{FS-BoW}), the 43 structural features (\textbf{Str}), 
and the best performing \textbf{CNN} classifier. As feature selection did not improve the performance of \textbf{Str} features, we do not show the performance of feature selection on \textbf{Str} in Table~\ref{tab:results}. As we can see from the table, the BoW-PD (Random Forest) is the highest performing model across all three datasets in terms of most compared measures.  

For example, on UNT.edu, BoW-PD achieves the highest precision of $0.90$ compared with all the other models, at the expense of a small drop in recall. Str performs better than BoW-all (Random Forest) and CNN in terms of recall and F1-score, i.e., Str achieves an F1 of $0.83$ as compared with $0.80$ and $0.79$ achived by BoW-all and CNN, respectively. Feature selection improves the performance of BoW-all classifier, i.e., FS-BoW (top-300 features) achieves an F1 of $0.85$ as compared with $0.80$ achieved by BoW-all. 

On Texas.gov, BoW-PD achieves a highest F1 of $0.78$ when we use the first-700 words combined with last-700 words from each document. Similar to UNT.edu, we can see the substantial improvement in the performance of BoW-all when we do feature selection, i.e., FS-BoW (top-2000 features) achieves an F1 of $0.71$ as compared with $0.66$ achieved by BoW-all. 
Moreover, the CNN classifier achieves the highest precision of $0.89$. 
CNN performs better than BoW-all, FS-BoW and Str, i.e., CNN achieves an F1 of $0.72$ as compared with $0.66$, $0.71$, and $0.64$ achieved by BoW-all, FS-BoW, and Str. 

On USDA.gov, BoW-PD achieves the highest score across all the measures when we use the first-2000 words, i.e., it achieves a highest F1 of $0.85$. Unlike the other two datasets, feature selection on the BoW did not improve the performance of the BoW, i.e., FS-BoW (top-1000 features) achieves an F1 of $0.79$ as compared with $0.80$ achieved by BoW-all. However, FS-BoW also achieves a highest recall of $0.88$ similar to BoW-PD. 

 \section{Conclusion and Future Directions}
\label{sec:conclusion}
In this paper, we studied different types of features and learning models to accurately distinguish documents of interest for a collection, from web archive data. Experimental results show that BoW features extracted using only some portions of the documents outperform BoW features extracted using the entire content of documents (full text) as well as the top-$N$ selected BoW features, structural features, top-$N$ selected structural features, and a CNN classifier. We found that feature selection done using information gain improved the performance of the BoW classifier. 
However, our conclusion is that text from specific portions of documents (e.g., the first-$X$ or first-$X$+last-$X$ number of words from the content of documents) is more useful than the text from the entire content for finding documents of interest to a given collection. 

Our aim was to provide interpretable models that are useful for librarians and collection managers 
to identify publications that match a given collection. 
Because of this, many traditional algorithms and approaches were selected 
so that we could easily interpret the output and could communicate what is happening within the models to librarians and collection creators who might not have as strong of an understanding of machine learning algorithms. In the future work, other approaches may produce more powerful models that could be more difficult to interpret or explain. Such models and in depth explorations of deep learning will be tested in the future. Moreover, it would be interesting to explore various dynamic model combinations that could select the most confident models for various categories of documents and that could improve the performance further. Our datasets and code will be made available to the research community to further research in this area. 

\section{Acknowledgements}
\label{sec:acknowledgements}
This work is generously funded by the Institute of Museum and Library Services (IMLS) under award LG-71-17-0202-17.


\bibliographystyle{ACM-Reference-Format}
\bibliography{sample-bibliography,hp_classi}

\end{document}